# Improved Riemannian Potato Field: an Automatic Artifact Rejection Method for EEG

Davoud Hajhassani[a,*], Quentin Barthélemy[b], Jérémie Mattout[c], and Marco Congedo[a]

[a] *GIPSA-Lab, University Grenoble Alpes, CNRS, Grenoble INP, Grenoble, France*

[b] *FOXSTREAM, Vaulx-en-Velin, France*

[c] *Lyon Neuroscience Research Centre, COPHY team, INSERM, University Claude Bernard Lyon1, Lyon, France*

[*] *Corresponding author at: GIPSA-Lab, University Grenoble Alpes, 11 Rue des Mathématiques, 38400 Saint-Martin d'Hères, France; e-mail:* [davoud.hajhassani@gipsa-lab.grenoble-inp.fr](davoud.hajhassani@gipsa-lab.grenoble-inp.fr) *(Davoud Hajhassani)*

ABSTRACT

Electroencephalography (EEG) signal cleaning has long been a critical challenge in the research community. The presence of artifacts can significantly degrade EEG data quality, complicating analysis and potentially leading to erroneous interpretations. While various artifact rejection methods have been proposed, the gold standard remains manual visual inspection by human experts—a process that is time-consuming, subjective, and impractical for large-scale EEG studies. Existing techniques are often hindered by a strong reliance on manual hyperparameter tuning, sensitivity to outliers, and high computational costs. In this paper, we introduce the improved Riemannian Potato Field (iRPF), a fast and fully automated method for EEG artifact rejection that addresses key limitations of current approaches. We evaluate iRPF against several state-of-the-art artifact rejection methods, using two publicly available EEG databases, labeled for various artifact types, comprising 226 EEG recordings. Our results demonstrate that iRPF outperforms all competitors across multiple metrics, with gains of up to 22% in recall, 102% in specificity, 54% in precision, and 24% in F1-score, compared to Isolation Forest, Autoreject, Riemannian Potato, and Riemannian Potato Field, respectively. Statistical analysis confirmed the significance of these improvements ($p < 0.001$) with large effect sizes (Cohen's $d > 0.8$) in most comparisons. Additionally, on a typical EEG recording iRPF performs artifact cleaning in under 8 milliseconds per epoch using a standard laptop, highlighting its efficiency for large-scale EEG data processing and real-time applications. iRPF offers a robust and data-driven artifact rejection solution for high-quality EEG pre-processing in brain-computer interfaces and clinical neuroimaging applications.

## 1. Introduction

Scalp electroencephalography (EEG) measures the post-synaptic potentials generated by neuronal activity through multiple electrodes positioned on the scalp [1,2]. Over almost a century, EEG has been widely employed as a neuroimaging modality both in clinical and cognitive neuroscience [3]. Thanks to its affordability, portability, non-invasiveness, and high temporal resolution, it has also become the modality of choice for non-invasive brain-computer interface (BCI) [4]. EEG features severe limitations though, for instance, poor spatial resolution and low signal-to-noise ratio, the latter mainly due to the fact that EEG signals are invariably contaminated by various types of electrical signals of non-cerebral origin, known collectively as artifacts. These are broadly categorized as either endogenous or exogenous [5–7].

Endogenous artifacts originate from biological sources picked up by scalp electrodes, including eye-related (eye blinks and eye movements), myogenic (facial/neck/shoulder muscle contraction, jaw, chewing, and hypoglossal (tongue) movements), heart-related (pulses time-locked to the QRS complex, cardioballistic) and sweat artifacts. Those are mostly involuntary and difficult to filter out as their spectrum largely overlaps with the spectrum of EEG [6]. Exogenous artifacts can be of environmental and instrumental origin. The former includes power-line (typically at 50Hz or 60Hz) and other electromagnetic interference. The latter includes electrode pop, disconnection, or malfunction and electrode impedance variations caused by the gradual drying of conductive gel [8]. While exogenous artifacts can be minimized in a controlled laboratory setting, they remain unavoidable in certain contexts, such as ambulatory EEG monitoring and real-world BCI applications [6].

Artifacts frequently exhibit considerable amplitude, which significantly degrades the signal-to-noise ratio, disrupts analysis, and renders many applications unfeasible or impractical. Due to the nonlinear characteristics of these artifacts, isolating them without compromising genuine neuronal data has proven to be a formidable task [9,10]. To make matters worse, in real-time applications such as BCI and neurofeedback, where system performance relies on the quality of the data, the presence of artifacts can disrupt the feedback stream [11,12]. These hurdles highlight the importance of including an artifact-handling stage in EEG signal analysis to effectively eliminate artifacts.

The gold-standard method for removing artifacts involves the visual inspection of raw EEG data by a human expert. This manual annotation procedure, however, is time-consuming and subjective. One contributing factor is the significant inter-subject variability in EEG data. Ideally, multiple researchers would process the same data, and their collective decisions would guide the data-cleaning process [5,13]. Because of the ever-growing large-scale EEG studies across different disciplines [14–16], manual removal of artifacts is becoming increasingly ineffective and impracticable. This has given rise to the development of several automated artifact removal techniques, which not only aim at saving time but also at allowing scalability and reproducibility. The artifact removal techniques developed so far fall into two main categories: (1) *artifact correction* techniques, which aim at modifying or adjusting the recorded signal to mitigate artifacts while preserving as much genuine EEG data as possible, and (2) *artifact rejection* techniques, which aim at identifying and discarding segments of data containing artifacts.

Different approaches have been introduced for the first category, ranging from Regression, Adaptive Filtering, Independent Component Analysis (ICA), Principal Component Analysis (PCA), and Canonical Correlation Analysis (CCA), to Empirical Mode Decomposition (EMD), Wavelet Transform (WT), and hybrid methods [6,8,10,17–19]. However, each of these methods comes with certain limitations, including the lack of full automation, reliance on additional reference channels, sensitivity to specific types of artifacts, performance dependency on the number of EEG channels, the need for expert intervention, and challenges in meeting real-time processing requirements [19]. The challenges specific to each approach, as summarized in Table 1, have driven the EEG research community to shift from single-method solutions to hybrid techniques, which have demonstrated superior performance [20,21]. However, these hybrid approaches—combining two or more individual methods—introduce increased computational complexity [22–24]. Recent research has increasingly focused on artifact correction techniques based on deep learning [25–27]. While these approaches often yield improved performance, they typically entail higher computational costs—making them unfit for real-time use—and require large amounts of training data, an aspect that remains challenging in the context of brain signals. Therefore, further investigation into transfer learning and data augmentation strategies for EEG data is essential to address the issue of data scarcity. Furthermore, these methods are often criticized for their lack of explainability. The main reason is that, unlike traditional algorithms, feature extraction in deep learning is opaque; therefore, it is difficult to appreciate the specific features influencing the decision [28,29]. They can also lead to poor generalization performance and scalability [30]. Additionally, it is essential to acknowledge that artifact correction methods may inadvertently modify valuable information in the EEG signal, potentially affecting subsequent analyses, particularly in cognitive and clinical EEG studies.

Table 1: Overview of the Challenges for Artifact Correction Methods

| Method | Reference Channel Required | Operator Expertise | Artifact-Specific Sensitivity | Fully Automated | Real-time Feasibility | Citations |
|---|---|---|---|---|---|---|
| Regression | Yes | Medium | High | Yes | Yes | [31,32] |
| Adaptive Filtering | Yes | Medium | High | Yes | Yes | [33,34] |
| ICA | Mostly Not | High | Low | Mostly Not | No | [35,36] |
| CCA | No | Low | Low | Mostly Yes | Yes | [37,38] |
| PCA | No | Low | Low | Mostly Yes | Yes | [39,40] |
| WT | No | Low | Low | Mostly Yes | No | [41,42] |
| EMD | No | Low | Low | Mostly Yes | No | [43,44] |
| Hybrid* | Mostly Not | Medium | Medium | Mostly Yes | Mostly Not | [41,45–47] |

*For hybrid methods, the specific characteristics may vary depending on the selected combination of individual techniques.

Given the limitations and challenges associated with artifact correction techniques, artifact rejection methods—which are also the primary focus of this work—offer a valuable and effective alternative. These methods eliminate the need for a reference channel, require minimal operator expertise, are applicable to all artifact types, and are predominantly automated. Importantly, they do not alter the EEG data in any way, thus establishing them as the standard for best research practices. Nevertheless, these approaches also present certain drawbacks, most notably data loss, which can be particularly problematic in scenarios where EEG data acquisition is costly or constrained. For this category, popular software tools such as Brainstorm [48], FieldTrip [49], EEGLAB [50], and MNE [51] provide the ability to identify and exclude data segments affected by artifacts. The general strategy of existing methods is to assess several data metrics, such as peak-to-peak amplitude, and reject data segments that exceed a manually set, fixed threshold value. Although this appears straightforward from a practitioner's perspective, it is not always practical. Due to the large variability of EEG data and, even more so, artifacts, a common problem they face is that the fixed threshold is sub-optimal. Instead, thresholds should be data-specific and automatically adapt to the data.

The necessity for enhanced automated artifact rejection methods is widely acknowledged by research teams, as evidenced by recent literature. For instance, pipeline-based approaches, such as FASTER (Fully Automated Statistical Thresholding for EEG Artifact Rejection), utilize fixed thresholds based on classical Gaussian statistics to identify electrodes and data segments affected by artifacts [52]. On the other hand, methods such as PREP [53] are specifically designed to detect and correct contaminated electrodes by interpolation, but do not offer a solution for eliminating data segments impacted by artifacts. However, most of these methods have hyperparameters that are dataset-dependent and require manual tuning. In this context, Autoreject (AR), an automated artifact rejection method for EEG data, has enjoyed considerable success and has been applied in various research studies. AR adaptively selects thresholds to distinguish between artifacts and clean data segments using cross-validation and employs Bayesian optimization to identify the optimal threshold [54]. However, its reliance on an interpolation algorithm and Bayesian optimization leads to slower performance when handling large databases, and makes it unsuitable for real-time applications. To address this issue, an automatic artifact rejection technique for extensive EEG recordings based on the Isolation Forest (IF) algorithm has been recently developed [55]. This approach eliminates the need for manually selecting a peak-to-peak amplitude threshold for artifact rejection by iteratively applying IF algorithm. The process continues until the distance metric between clean data segments and artifacts remains unchanged. It has also been shown that IF outperforms AR and deep learning methods such as EEGdenoiseNet [56] in terms of overall data quality and execution time.

Despite its relatively short history, the application of Riemannian geometry in EEG signal processing is gaining significant attention due to its simplicity, robustness, outstanding performance in EEG-based BCI decoding, and capabilities in transfer learning [57]. This approach has led to the introduction of the Riemannian Potato (RP), a multivariate artifact rejection technique that considers any EEG signal deviating significantly from the normal baseline signal as an artifact [58]. The key feature of interest is the covariance matrix derived from EEG epochs. RP estimates the barycenter of all covariance matrices and evaluates the distance of each from the barycenter by employing appropriately derived z-scores. Eventually, RP discards epochs whose covariance matrices fall outside an acceptance region set by a predefined z-score threshold. The name of the method is inspired by the *potato*-like shape of the z-scores' isocontour, determined by the non-linearity of the Riemannian manifold, as illustrated in Fig. 1. The RP method has been extensively employed for online artifact rejection in P300-based BCI spellers [59] and games [60], for offline rejection prior to the statistical analysis of cognitive assessments [61], and in the detection of epilepsy [62]. A major drawback of RP is decreased sensitivity and specificity as the number of electrodes increases.

Several enhancements have been proposed for RP, including the incorporation of robust barycenter estimation by removing outliers [63] and the use of geometric statistics instead of arithmetic statistics to derive z-scores [57,64]. The Riemannian Potato Field (RPF) is the most recent development, serving as a generalization and extension of RP [65]. In addition to employing a robust barycenter estimator, it addresses the decline in performance caused by an increasing number of electrodes by utilizing multiple low-dimensional potatoes in parallel, each tailored to detect specific types of artifacts that influence particular spatial regions within certain frequency bands. The output z-scores from all potatoes (*i.e.*, a potato field) are merged into a single p-value using Fisher's combination function [66], allowing a Signal Quality Index (SQI) for each epoch ranging from 0 (noisy signal) to 1 (clean signal). Eventually, RPF discards epochs with an SQI falling below a predefined threshold. Despite these advancements, Riemannian-based artifact rejection techniques still rely on manually defined fixed thresholds, which limits their adaptability across different datasets and constrains their automation. Moreover, neither the RP nor RPF methods provide a solution for handling EEG data with a high number of outliers, which is particularly common in clinical EEG recordings. In these methods, if the proportion of outliers within the EEG data is substantial, the barycenter estimation may still be influenced by them, even when using a robust barycenter estimator.

In this work, we present an improved Riemannian Potato Field (iRPF), an automatic and adaptive artifact rejection technique designed to overcome key limitations of the RP and RPF methods. The iRPF method adaptively determines all required thresholds by using a knee detection algorithm [67]. Additionally, iRPF addresses sensitivity to outliers by integrating an automatic outlier rejection mechanism that sets an adaptive threshold based on the field root mean square (FRMS) of the EEG data, ensuring effective barycenter estimations even in the presence of extreme outliers. Furthermore, iRPF expands the distance metrics used in artifact detection by incorporating the Euclidean and diagonal-Euclidean distance, which are particularly effective in detecting vertical eye movements and myogenic artifacts. Finally, iRPF integrates multiple p-value combination functions alongside Fisher's combination function and introduces a meta-combination strategy to aggregate p-values, leading to more accurate SQI computations. The Julia and Python implementations of the iRPF method are made available in a publicly accessible GitHub repository [68].

Our preliminary work [69] introduced the initial concepts of the iRPF within an EEG-based BCI pipeline to explore its practical potential in the context of BCIs. In the present study, we significantly extend this work by systematically evaluating the effectiveness of iRPF in maximizing artifact rejection and preserving clean neural data. Specifically, we compare its performance in terms of recall, specificity, precision, F1-score, and execution time against four state-of-the-art methods (RPF, RP, AR, and IF) using two publicly available EEG databases encompassing various artifact types.

The remainder of the paper is organized as follows: The Materials and Methods section offers an overview of essential concepts in Riemannian geometry, along with details on RP and RPF. Following this, the proposed iRPF method is presented. The Results section presents findings derived from comparisons between iRPF and the selected state-of-the-art artifact rejection methods. The Discussion section considers the advantages and limitations of this work, along with several possible directions for further research on artifact rejection methods.

## 2. Methods

In this section we denote matrices by upper case bold letters (**A**), variables by lower case italic letters (*a*), and constants by upper case italic letters (*A*). The matrix operators $(.)^T$, $(.)^{-1/2}$, $Log()$, and $\|.\|_F$ denote the transpose, inverse of the principal square root, logarithm, and Frobenius norm of the argument, respectively.

### 2.1. Geometry of Covariance Matrices

Hereafter $\mathbf{X} \in \mathbb{R}^{N \times T}$ represents an EEG epoch, recorded across $N$ channels/electrodes at $T$ temporal samples. If the signal is centered, such as after applying band-pass filtering, the maximum-likelihood estimator of the covariance matrix of **X** is

$$\mathbf{\Sigma} = \frac{1}{N-1} \mathbf{X}\mathbf{X}^T \in \mathbb{R}^{N \times N}. \qquad (1)$$

The efficiency of this estimator requires $T$ to significantly exceed $N$ [70]. For comparison with other estimators, the interested reader is directed to [71]. Covariance matrices are symmetric positive definite (SPD), thus they are processed within a Riemannian manifold $\mathcal{M}_N$, of dimension $m = (N(N+1)/2)$, possessing a distinct geometry [64,72,73]. Employing the eigenvalue decomposition of $\mathbf{\Sigma}$,

$$\mathbf{\Sigma} = \mathbf{U} \operatorname{diag}(\lambda_1, \ldots, \lambda_N)\mathbf{U}^T,$$

where $\lambda_1, \ldots, \lambda_N$ are the eigenvalues of $\mathbf{\Sigma}$ and **U** is the matrix holding in the columns the corresponding eigenvectors, the unique symmetric inverse square root $\mathbf{\Sigma}^{-\frac{1}{2}}$ is computed as

$$\mathbf{\Sigma}^{-\frac{1}{2}} = \mathbf{U} \operatorname{diag}\left(\lambda_1^{-\frac{1}{2}}, \ldots, \lambda_N^{-\frac{1}{2}}\right)\mathbf{U}^T.$$

The affine-invariant Riemannian distance between two points $\mathbf{\Sigma}_1$ and $\mathbf{\Sigma}_2$ is given by [74]

$$\delta_R(\mathbf{\Sigma}_1, \mathbf{\Sigma}_2) = \left\| Log\left(\mathbf{\Sigma}_1^{-\frac{1}{2}} \mathbf{\Sigma}_2 \mathbf{\Sigma}_1^{-\frac{1}{2}}\right)\right\|_F = \left(\sum_{n=1}^N log^2 \ell_n\right)^{\frac{1}{2}}, \qquad (2)$$

where $\ell_n, n = 1, \ldots, N$, are the eigenvalues of $\mathbf{\Sigma}_1^{-\frac{1}{2}} \mathbf{\Sigma}_2 \mathbf{\Sigma}_1^{-\frac{1}{2}}$ or $\mathbf{\Sigma}_1^{-1}\mathbf{\Sigma}_2$. The barycenter, or geometric mean, of $I$ matrices $\mathbf{\Sigma}_i$, where $i = 1, \ldots, I$, is defined as the matrix that minimizes the dispersion such as

$$\overline{\mathbf{\Sigma}} = \arg \min_{\mathbf{\Sigma} \in \mathcal{M}_N} \sum_{i=1}^I \delta_R^2(\mathbf{\Sigma}_i, \mathbf{\Sigma}). \qquad (3)$$

The Riemannian mean is more resilient to outliers as compared to the Euclidean mean [57]. However, it is adversely affected by ill-conditioned input matrices. The estimation requires iterative algorithms such as the gradient descent algorithms [75]. In practice, the GM-GD algorithm proposed in [76] or the fixed-point algorithm in [77], both implemented in the *pyRiemann* Python package, are those most commonly used [78].

### 2.2. Riemannian Potato

The advantage of the RP method lies in its ability to utilize multivariate features for artifact detection [58]. This approach involves representing EEG epochs as covariance matrices, estimating a barycenter $\overline{\mathbf{\Sigma}}$ and obtaining a z-score to quantify the distance of each covariance matrix from the barycenter (Fig. 1). As illustrated in Fig. 2, artifacts determine the peculiar structure of the covariance matrix and significantly influence its distance from the barycenter estimated on clean epochs. The implementation (Algorithm 1) is straightforward: for each epoch indexed by $i$, the affine-invariant Riemannian distance between the $i$th covariance matrix $\mathbf{\Sigma}_i$ and the barycenter $\overline{\mathbf{\Sigma}}$ is computed as $d_i = \delta_R(\mathbf{\Sigma}_i, \overline{\mathbf{\Sigma}})$. Then, the z-score $z_i$ of this distance is computed as

$$z_i = \frac{d_i - \mu}{\sigma}, \qquad (4)$$

where $\mu$ and $\sigma$ are the arithmetic mean and standard deviation of the distances to the barycenter. The z-score is used to quantify the dispersion of covariance matrices classified as clean, forming a SQI and enabling the determination of an appropriate threshold $z_{th}$ (typically equal to 2.0) to define the rejection region. As discussed and as we will show, due to inter-subject variability in EEG and artifact variability, a prior fixed threshold is sub-optimal. Since RP lacks a dedicated outlier rejection step, the Riemannian barycenter may be influenced by outliers, leading to the inclusion of artifacts in the barycenter estimation and ultimately hindering rejection performance.

---
**Algorithm 1: Riemannian Potato**

**Initialization:** $z_{th} = 2.0$, **Input:** $\mathbf{\Sigma}_i$, $i \in \{1, \dots, I\}$

---
**1:** Estimation of the barycenter $\overline{\mathbf{\Sigma}}$ from the set of covariance matrices $\{\mathbf{\Sigma}_i\}_{i=1}^{I}$ with Eq. (3)
**2:** Estimation of the $I$ distances $d_i$ to the barycenter $\overline{\mathbf{\Sigma}}$ with Eq. (2)
**3:** Estimation of the mean $\mu$ and the standard deviation $\sigma$ of $I$ distances
**4:** Computation of z-score $z_i$ (SQI) with Eq. (4)
**5:** **if** $z_i > z_{th}$ **then** Epoch $i$ is considered as artifacts
   **else** Epoch $i$ is considered as clean data

---

*2.3. Riemannian Potato Field*

The RP method was further developed and enhanced in [65]. RP is typically applied across all $N$ channels. However, due to the multivariate nature of such a potato, as shown in Fig. 2, artifacts that cause significant variations in only one or a few channels may not determine a substantial deviation of the overall covariance matrix from the barycenter. Consequently, such artifacts may remain undetected. It has been observed that this phenomenon can occur for a headset with as few as $N = 8$ channels [65]. To overcome this limitation, the RPF incorporates and integrates multiple potatoes of different dimensions. Each of these potatoes is designed to target specific artifacts that commonly affect particular spatial regions (*e.g.*, subsets of channels) in specific frequency bands. For example, to target eye-related artifacts, a potato is defined using EOG and/or forehead electrodes, with the signal low-pass filtered below 7 Hz. In the case of myogenic artifacts, multiple potatoes can be defined using external electrodes and a high-pass filter above 20 Hz. This may include a potato utilizing temporal electrodes to detect jaw clenching and swallowing. Ultimately, the z-scores obtained from all potatoes are subsequently combined into a unified p-value, which serves as a SQI.

After defining a set of $J$ potatoes forming the RPF, their output z-scores are combined into a single p-value using the Fisher's combination function [66]. For z-scores $z_j$, $j = 1, \dots, J$, their corresponding p-values $p_j$ are obtained by $p_j = 1 - \text{cdf}_{N_{0,1}}(z_j)$, where $\text{cdf}_D$ represents the cumulative distribution function of $D$, and $N_{0,1}$ denotes the normal distribution with mean 0 and standard deviation 1. The Fisher's function combines the p-values $p_j$ as

$$q = -2 \sum_{j=1}^{J} \log(p_j). \qquad (5)$$

Under the assumption of uniformly distributed and independent p-values, $q$ is distributed as a chi-square with $2J$ degrees of freedom. Thus, its p-value is obtained as

$$p = 1 - \text{cdf}_{\chi^2_{2J}}(q).$$

In the RPF, $p \in [0,1]$ is considered as a SQI which can be employed for quality assessment and artifact rejection. In the case of artifact rejection, a threshold $p_{th}$, which is typically set to 0.01, is applied to p-values. Consequently, EEG epochs are classified as artifacts whenever their corresponding $p$ falls below the threshold. However, the z-scores of affine-invariant Riemannian distances are not normally distributed, which means that $p$ is generally not uniformly distributed. This implies that the threshold is not meaningful in absolute terms. Once again, while it is possible to empirically set an effective threshold, the optimal value strongly depends on the data, to which the threshold should adapt. This represents a major limitation of the RP and RPF methods, which will be addressed in the following section.

The RPF method also employs a robust barycenter estimator by offline outlier removal [63]. In fact, although the Riemannian barycenter demonstrates robustness to outliers, it can nonetheless be affected by them [57]. The robust estimator computes the barycenter iteratively in a fixed number of steps (typically, three to five) by excluding at each step all covariance matrices with a z-score superior to a pre-defined threshold (typically equal to 2.0) [63,71]. However, the pre-defined threshold is not optimal, as it remains constant for all potatoes and across different subjects. The iRPF method presented in the following section eliminates the need to tune this hyperparameter as well.

The SQI in RP is computed using arithmetic statistics, which are suboptimal due to the non-normal distribution of affine-invariant Riemannian distances. Since these distances empirically display a highly right-skewed positive-only distribution, they are better modeled by a log-normal or chi-squared distribution. As a result, geometric statistics provide more effective descriptors [57,64]:

$$\mu = exp\left(\frac{1}{I}\sum_{i=1}^{I} \log(d_i)\right),$$

$$\sigma = exp\left(\sqrt{\frac{1}{I}\sum_{i=1}^{I}(\log(d_i/\mu))^2}\right),$$

$$z_i = \frac{\log(d_i/\mu)}{\log(\sigma)}.$$

These geometric statistics have been used in several offline RP methods [59,71], in RPF, and are maintained in our proposed methods.

*2.4. Improved Riemannian Potato Field*

The proposed iRPF method introduces several key improvements over the RPF [65] to address the limitations of Riemannian-based artifact rejection methods, which will be discussed in the following sections.

*2.4.1 Automatic outlier rejection*

In this study, we present a novel method for automatically rejecting large outliers from the EEG recording, ensuring that barycenter estimations are not biased by extreme values— addressing a limitation of both the RP and RPF methods. This process begins with the computation of the FRMS across the entire EEG recording. The FRMS is defined as

$$\text{FRMS}(t) = \sqrt{\frac{1}{N}\sum_{n=1}^{N} x(n,t)^2}, \quad t = 1, \dots, T$$

where, $x(n,t)$ denotes the EEG signal at electrode $n$ and time $t$. The FRMS corresponds to the square root of the mean global field power (GFP), a well-established measure of global EEG power as a function of time that has been widely used in EEG research [79]. The GFP represents the overall amplitude of each time sample, providing a measure by incorporating data from all recording electrodes simultaneously [80].

Next, as shown in Fig. 3, the mean FRMS value $\mu_{frms}$ is computed within a window twice the window-length, centered around the median of the sorted FRMS values. This serves as a median estimator. To prevent samples with no signal (*i.e.,* zero everywhere) from influencing the threshold, the lower limit $l_{lim}$ is set at the first non-zero element in the sorted FRMS values. The outlier rejection threshold $th_{rej}$ is then defined as

$$th_{rej} = \mu_{frms} + u_{lim}(\mu_{frms} - l_{lim}),$$

where $u_{lim}$ is the upper limit, set to 1 by default but adjustable based on the characteristics of the database. The underlying assumption is that, in normal EEG, the distribution of FRMS values is expected to be finite and symmetric. Therefore, the absolute FRMS values are expected not to significantly exceed twice the distance $\mu_{frms} - l_{lim}$. Eventually, epochs are considered outliers and rejected if any of their samples fall above the threshold $th_{rej}$. This corresponds to amplitude thresholding done ubiquitously, for example in the *MNE* Python package [51], but in an adaptive fashion.

*2.4.2 Adaptive robust barycenter estimation*

The robust barycenter estimations are achieved using an adaptive algorithm that eliminates the need for manually setting an outlier rejection threshold. At each iteration, the algorithm calculates the geometric z-scores based on the distance of each point from the current barycenter estimation. These z-scores are then transformed into p-values. Once the p-values are calculated, they are arranged in ascending order to identify potential outliers. To determine how many of these points should be discarded in the subsequent iterations, the iRPF method applies a knee-detection algorithm known as *Kneedle* [67]. This technique identifies the point in the ordered list where the rate of change in the p-values increases sharply (the "knee"), suggesting a boundary between inliers and outliers. The algorithm halts when no further knee is detected in the sorted p-values, or after a maximum of four iterations, a limit determined through preliminary experiments and prior research [81]. These studies showed that additional iterations do not significantly enhance the robustness of the barycenter estimation and only increase computational cost. This approach allows the estimation to adapt dynamically to the data distribution, providing a more resilient barycenter without the need for a fixed threshold.

*2.4.3 More distance feature to build potato fields*

The RP and RPF methods rely on the affine-invariant Riemannian distance between covariance matrices and the barycenter. In this study, we expand upon this framework by introducing additional distance features to capture a broader range of artifact types, specifically incorporating the Euclidean distance and the diagonal Euclidean distance. The Euclidean distance between two connivance matrices $\boldsymbol{\Sigma}_1$ and $\boldsymbol{\Sigma}_2$ is given by

$$\delta_E(\boldsymbol{\Sigma}_1, \boldsymbol{\Sigma}_2) = \|\boldsymbol{\Sigma}_1 - \boldsymbol{\Sigma}_2\|_F.$$

This distance is particularly advantageous for detecting artifacts that are expected to co-vary across electrodes, irrespective of the sign of the co-variation. For instance, as illustrated in Fig. 4, vertical eye movements affecting the contra-lateral frontal electrodes can be more effectively identified by computing the Euclidean distance between each covariance and the barycenter.

The diagonal Euclidean distance between two points $\boldsymbol{\Sigma}_1$ and $\boldsymbol{\Sigma}_2$ is defined as

$$\delta_{diag(E)}(\boldsymbol{\Sigma}_1, \boldsymbol{\Sigma}_2) = \|diag(\boldsymbol{\Sigma}_1 - \boldsymbol{\Sigma}_2)\|_F,$$

where $diag$ is the operator nullifying the off-diagonal elements of the argument. This distance is particularly effective for identifying artifacts that do not exhibit co-variation across electrodes, such as myogenic artifacts affecting the peripheral electrodes, which primarily impact the diagonal elements of the covariance matrices (as shown in Fig. 2). Following the same approach as with the affine-invariant Riemannian distance, we compute the z-scores of these new distance features, transform them into p-values, and subsequently combine them to derive the SQI.

*2.4.4 Various combination and meta-combination functions*

The RPF uses the Fisher combination function Eq. (5) to combine the p-values obtained for each potato, while the literature presents a diverse range of p-value combination functions, each exhibiting distinct statistical properties. In this study, we introduce the integration of several additional p-value combination functions, specifically the Liptak, Pearson, and Tippett functions [82], each of which possesses distinct sensitivity characteristics. The Fisher combination function is sensitive to the presence of small p-values, while the Pearson combination function is sensitive to the presence of large p-values. The Tippett combination function focuses exclusively on the smallest p-value, while the Liptak combination function weights in a fair manner all p-values. Once the p-values corresponding to $J$ z-scores have been obtained, the Liptak function combines the p-values $p_j$ as

$$q = \frac{1}{J}\sum_{i=1}^{J} \Phi^{-1}(p_j),$$

where $\Phi$ denotes the standard normal distribution cumulative function. Under the assumption of uniformly distributed and independence, the combined p-value is then obtained as $p = 1 - \text{cdf}_{N_{0,1}}(q)$. For Pearson's function, the combined p-value is calculated using

$$q = -2\sum_{j=1}^{J} \log(1 - p_j).$$

Under the assumption of uniformly distributed and independence, $q$ follows a chi-square distribution with $2J$ degrees of freedom, leading to the combined p-value defined as $p = \text{cdf}_{\chi^2_{2J}}(q)$. In contrast, the Tippett function combines the p-values as

$$q = \min(p_1, \ldots, p_J),$$

with the corresponding p-value given by $p = \text{cdf}_{\beta_{1,J}}$, where $\beta_{1,J}$ denotes the Beta distribution with shape parameters 1 and $J$.

We also introduce the use of meta-combination functions, which combine multiple combination functions. For instance, in this study, we applied the Tippet combination function to the results of the Liptak and Fisher combination functions. This approach enhances the sensitivity of SQI computation by leveraging the strengths of multiple combination functions, resulting in a more precise determination of the rejection region, as shown in Fig. 5.

*2.4.5 Automatic and adaptive thresholding for artifact rejection*

The final improvement introduced by the iRPF method is the automatic and adaptive determination of the threshold for rejecting artifacts. In contrast to the RPF method, where this threshold is manually pre-defined, the iRPF employs an adaptive approach using the knee-detection algorithm [67]. This algorithm dynamically adjusts to the characteristics of the data, identifying the optimal threshold (the "knee") based on the sorted SQI values. This eliminate the need for manually setting the rejection region. However, the effectiveness of the knee detection algorithm largely depends on the distribution of the SQIs. As illustrated in Fig. 6, the knee detection algorithm performs more effectively in defining a rejection region in the iRPF method, where p-values are derived using additional distance metrics alongside the affine-invariant Riemannian distance and are combined into SQIs through a meta-combination function, as compared to the RPF method.

*2.5. Design of the Potato Field*

Since each potato in the field is designed for specific spatial regions and frequency bands, the potato field in iRPF, like in RPF, must be customized for each EEG headset, considering the unique localization of its channels. We exemplify the design of the potato field for a headset with $N = 21$ EEG electrodes, positioned according to the international 10-10 system [83], which includes pre-frontal (Fp1, Fpz, Fp2), frontal (F7, F3, Fz, F4, F8), temporal (T7, T8), central (C3, Cz, C4), parietal (P7, P3, Pz, P4, P8), and occipital (O1, Oz, O2) electrodes. To illustrate the utilization of EOG channels, we consider six additional electrodes positioned at the outer canthus, superior, and inferior of the left eye (EOGL1, EOGL2, EOGL3) and the right eye (EOGL1, EOGL2, EOGR3) [84].

### 2.5.1 Potato for eye-related artifacts

Eye-related artifacts generally appear as high-amplitude variations in brain signals due to eye blinks or as low-frequency patterns caused by horizontal eye movements (HEM) and vertical eye movements (VEM), primarily contaminating the pre-frontal and frontal electrodes with frequencies inferior to 8 Hz [85,86]. To detect eye-related artifacts, the potato field is defined as shown in Table 2, whether EOG channels are available or not.

Table 2: Potato Field for Eye-related Artifacts

| Channels | Artifact | Distance | Frequency band (Hz) |
|---|---|---|---|
| If EOG channels are available | | | |
| EOGL1 EOGR1 | HEM | Riemannian | 0.1-7 |
| EOGL2 EOGL3 | VEM | Euclidean | 0.1-7 |
| EOGR2 EOGR3 | VEM | Euclidean | 0.1-7 |
| EOGL2 EOGL3 | Blink | Riemannian | 0.1-7 |
| EOGR2 EOGR3 | Blink | Riemannian | 0.1-7 |
| If EOG channels are not available | | | |
| Fp1 Fp2 | HEM | Riemannian | 0.1-7 |
| Fp1 Fp2 | VEM | Euclidean | 0.1-7 |
| Fp1 Fpz Fp2 | Blink | Riemannian | 0.1-7 |

### 2.5.2 Potato for myogenic artifacts

Myogenic artifacts exhibit a wide frequency range, typically exceeding 20 Hz [87]. Because these activities primarily affect the electrodes closest to the muscles, peripheral electrodes are employed in the design of the potato field as show in Table 3.

Table 3: Potato Field for Myogenic Artifacts

| Channels | Artifact | Distance | Frequency band (Hz) |
|---|---|---|---|
| F7 F8 | EMG | diag(Euclidean) | >20 |
| T7 T8 | EMG | diag(Euclidean) | >20 |
| P7 P8 | EMG | diag(Euclidean) | >20 |
| O1 Oz O2 | EMG | diag(Euclidean) | >20 |

### 2.5.3 Potato for electrode pop and dysconnectivity

These artifacts are generated by a temporary contact loss. They can occur at any electrode and often leads to a discontinuity in the recording, followed by a typical oscillatory period produced by the impulse response of various analog and digital filtering processes, ultimately returning to normal functioning. This phenomenon causes a sudden decrease between the covariance of the disconnected channels and all the others. As a result, any pairs of electrodes that are band-pass filtered between 1 and 20 Hz would effectively identify these events, using the affine-invariant Riemannian distance [65].

### 2.5.4 Potato for general artifacts

A variety of other artifacts can contaminate EEG signals, impacting channels and frequencies more or less specifically, such as cardiac and movement artifacts. To detect general artifacts, all channels are filtered in a large frequency band. Given that BCI or neurofeedback applications focus on modulating specific brain oscillations, it is advisable to exclude these frequencies from the overall potato design. In this case, the affine-invariant Riemannian distance can effectively identify these events [65].

### 2.6. Description of Databases

For the evaluation of the methods, we utilized two publicly available EEG databases: the EEG eye artifact (EYEAR) database [84,88,89] and the Temple University EEG Artifact (TUAR) corpus [90,91], the principal characteristics of which are outlined in Table 4.

Table 4: Principal Characteristics of Databases Used in This Study

| Database | Subjects (EEG files) | Sampling rate (Hz) | EEG channels | EOG channels |
| --- | --- | --- | --- | --- |
| EYEAR - study 01 | 5 | 200 | 58 | 6 |
| EYEAR - study 02 | 15 | 200 | 64 | 6 |
| EYEAR - study 03 | 10 | 200 | 64 | 6 |
| TUAR | 196 | 125 | 22 | 0 |

*2.6.1  EYEAR database*

The database comprises EEG data collected from five sub-studies. For our analysis, we specifically focused on the first three studies within the database [84,88,89], comprising a total of 30 participants. The database includes trials of resting state, HEM, VEM, and eye blinks. Given the variety of eye movement types recorded, this database is well-suited for evaluating the effectiveness of our method in detecting the aforementioned eye-related artifacts, and the ability to preserve clean brain signals. While each sub-study includes its own set of tasks, the owners only provide signals from the calibration phase, where participants perform eye movements. The three selected studies were recorded at a sampling rate of 200 Hz, though they employed two different electrode configurations. The database recordings had already undergone initial pre-processing, including notch filtering with cut-off frequencies at 49 and 51 Hz, and a 0.1 Hz high-pass filter using a 2nd-order Butterworth filter. Visual inspection was also conducted to exclude trials with myogenic artifacts or electrode shifting [92].

A careful visual inspection of the data revealed that some epochs labeled as "rest" were actually contaminated by artifacts, particularly eye blinks. Consequently, we manually reviewed the data and relabeled the affected epochs to ensure proper labeling. We divided each recording into non-overlapping 4-second EEG segments. The labels, originally assigned to 8-second intervals, were adjusted to correspond to the 4-second segmentation. Finally, raw epochs were band-pass filtered in the 0.1-44 Hz region using a 4th-order Butterworth filter. Since the potato field design is consistent across all studies in this database, we have combined all studies for analysis.

*2.6.2  TUAR Corpus*

The database represents a subset of the world's largest open-source clinical EEG corpus [16]. It comprises 310 EEG recordings from 213 patients, each meticulously annotated by experts to identify artifacts. Since the database includes real-world clinical data, it offers a diverse and challenging set of artifacts, including chewing, electrode, eye movement, muscle, and shiver, making it well-suited for evaluating the effectiveness of our method in detecting real clinical artifacts. The database includes annotations for the aforementioned artifacts, seven combinations of two different artifacts, and background EEG (no artifact). In this work, data segments containing any kind of artifacts are considered as the artifact class, with our goal being to distinguish them from clean background EEG.

After a comprehensive visual inspection of all EEG files, 196 recordings were selected for analysis, as they contained sufficient clean data necessary for training the potatoes. Files that were excluded exhibited highly contaminated backgrounds that did not represent clean data or contained artifacts throughout the entire recording. Subsequently, standard pre-processing steps were applied. First, all EEG recordings were down-sampled to 125 Hz. Next, the raw data were filtered using a band-pass filter in the range of 0.5–60 Hz and a notch filter with cut-off frequencies at 59 and 60 Hz, using a 4th-order Butterworth filter. Finally, the data were segmented into non-overlapping 5-second intervals, with segments labeled as either artifact or background EEG.

*2.7. Pipeline*

All data processing and analyses were performed using Julia (release version 1.11.2) and Python (release version 3.9.18) programming languages on a laptop equipped with Windows 10, an Intel i9-11950H @2.6 GHz CPU and 128 GB of RAM. The proposed iRPF has been implemented in Julia and then exported onto the Python environment using *JuliaCall* python package [93]. The minimal pre-processing of the raw EEG data and segmentation into epochs as detailed in the previous section, was performed using the *MNE* python package [51].

The performance metrics utilized in this study include recall, specificity, precision, and F1-score, and derived from the confusion matrix [94]. These metrics provide a comprehensive evaluation of the model's ability to distinguish between clean and artifact epochs. Recall measures the proportion of artifacts that are correctly detected and rejected, while specificity quantifies the proportion of clean epochs that are correctly identified and retained. To further assess the method's capability in artifact rejection and preservation of clean segments, precision is reported, which indicates the proportion of detected artifacts that were actually labeled as such. Additionally, the F1-score is employed to provide a balanced evaluation between precision and recall.

## 3. Results

*3.1. Comparison with the State of the Art*

In this subsection, we compare the proposed iRPF method with several state-of-the-art artifact rejection techniques, including both automatic and manual approaches. The first method considered for comparison is the RPF, which requires manually setting a rejection threshold for the SQI of all epochs [65]. For this evaluation, we set the threshold at 0.5, which yielded the highest averaged F1-score. Although the potato fields for RPF are defined identically to iRPF, RPF only uses the affine-invariant Riemannian distance, as it is the only available distance metric. The employed potato fields are detailed in Table 5. The second method is the RP, which also requires manually setting a rejection threshold for the z-scores of all epochs [58]. In this study, the threshold is set to 2.0, same with most previous studies. Since RP considers all channels collectively, there is no need to define a potato field. Both methods are available in the *pyRiemann* library [78]. To ensure a fair comparison, we applied the proposed automatic outlier rejection method (section 2.4.1) before using the RPF and RP. The third method is AR [54]. Although the AR method is capable of correcting EEG segments, for the sake of a fair comparison, its parameters were set to reject EEG segments if even a single channel was contaminated by artifacts. The last method is based on the IF algorithm [55]. It iteratively detects outliers in the EEG data and adjusts the boundary of inliers using a statistical indicator.

Table 5: Designed Potato Field

| **Database**: EYEAR Database – With EOG Channels | | |
|---|---|---|
| Potatoes | Distance | Frequency band (Hz) |
| EOGL2 EOGL3 | Euclidean | 0.1-7 |
| EOGR3 EOGR2 | Euclidean | 0.1-7 |
| EOGL2 EOGL3* | Riemannian | 0.1-7 |
| EOGR3 EOGR2* | Riemannian | 0.1-7 |
| EOGL1 EOGR1* | Riemannian | 0.1-7 |
| **Database**: EYEAR Database – Without EOG Channels | | |
| AF7 AF8 | Euclidean | 0.1-7 |
| AF7 AF8* | Riemannian | 0.1-7 |
| F7 F8 | Euclidean | 0.1-7 |
| F7 F8* | Riemannian | 0.1-7 |
| AF3 AF4 Fpz | Euclidean | 0.1-7 |
| AF3 AF4 Fpz* | Riemannian | 0.1-7 |
| F1 Fz F2* | Riemannian | 0.1-7 |
| **Database**: TUAR Database | | |
| Fp1 Fp2* | Riemannian | 1-5 |
| Fp1 Fp2 | Euclidean | 1-5 |
| F7 F8* | Riemannian | 1-5 |
| F7 F8* | diag(Euclidean) | 20-60 |
| T3 T4* | diag(Euclidean) | 20-60 |
| O1 O2* | diag(Euclidean) | 20-60 |

A * symbol indicates the potato fields designed for the RPF method, all utilizing the affine-invariant Riemannian distance.

Fig. 7 shows the evaluation metrics for the two databases, averaged across all subjects, for all considered artifact rejection methods, along with the corresponding normalized confusion matrices. Since many databases do not include EOG channels, the comparison for the EYEAR database was conducted with and without the use of EOG channels. As illustrated, the iRPF method consistently outperforms the other techniques across all performance metrics. Notably, iRPF achieves an optimal balance between recall and specificity, reflecting its ability to effectively remove artifacts while preserving clean data. In contrast, methods such as RP, IF, and AR exhibit lower specificity, indicating a tendency to mistakenly classify clean signals as artifacts.

To assess the statistical significance of the observed improvements, the iRPF method was tested against all methods chosen for comparison using all metrics as dependent variables. For each database considered, this yielded 16 paired t-tests. Significance was obtained by means of a multiple comparison permutation t-max step-down procedure using 200K random permutations. This procedure ensures that the family-wise error rate, *i.e.*, the probability of falsely rejecting one or more of the 16 null-hypotheses, does not exceed 0.05 [95]. The tests were carried out using the Julia package *PermutationTests.jl* [96]. Additionally, Cohen's d was computed to assess the effect sizes of the observed differences [97]. The results are presented in Table 6. For the EYEAR database, both with and without EOG channels, the iRPF method consistently outperformed the other methods, with large effect sizes observed in most comparisons. In the TUAR database, while the effect sizes were generally smaller, the iRPF method still demonstrated a significant advantage over several other methods and performance metrics.

Table 6: Effect Size and Statistical Tests for Assessing the Performance of iRPF Against the Four Chosen Competitors.

| Database | Performance Metric | | | | | | | | | | | | | | | |
|---|---|---|---|---|---|---|---|---|---|---|---|---|---|---|---|---|
| | Recall | | | | Specificity | | | | Precision | | | | F1-score | | | |
| | RPF | RP | IF | AR | RPF | RP | IF | AR | RPF | RP | IF | AR | RPF | RP | IF | AR |
| EYEAR with EOG | ** | ** | ** | | ** | ** | ** | ** | ** | ** | ** | ** | ** | ** | ** | ** |
| | 1.64 | 4.21 | 1.4 | 0.39 | 1.38 | 12.3 | 2.81 | 2.9 | 1.52 | 6.21 | 2.93 | 2.67 | 1.92 | 6.76 | 2.4 | 2.03 |
| EYEAR w/o EOG | ** | ** | * | | * | ** | ** | ** | ** | ** | ** | ** | ** | ** | ** | ** |
| | 1.3 | 2.46 | 0.71 | 0.18 | 0.81 | 11.7 | 2.62 | 2.54 | 0.95 | 5.55 | 2.85 | 2.39 | 1.5 | 4.41 | 2.05 | 1.42 |
| TUAR | ** | ** | ** | * | ** | ** | ** | ** | ** | ** | ** | ** | ** | ** | ** | |
| | 0.22 | 1.37 | 0.16 | 0.22 | 1.5 | 3.29 | 1.04 | 0.52 | 0.6 | 1.19 | 0.48 | 0.15 | 0.6 | 1.19 | 0.39 | 0.01 |

The table reports the Cohen's d effect sizes. For interpretation, the effect size can be considered as very small ($< 0.2$), small ($< 0.5$), medium ($< 0.8$), or large ($> 0.8$).
Legend: * = $p<0.05$, ** = $p<0.001$ (significance level of the multiple comparison permutation tests)

*3.2. The Importance of Automatic Thresholding*

In this subsection, the proposed iRPF method is thoroughly compared with its predecessor, the RPF method. To highlight the advantages of the iRPF method, particularly its adaptive automatic thresholding capability, different rejection thresholds for the RPF method were manually defined and applied across all subjects in the EYEAR database with and without using EOG channels. The definition of the potato field remains consistent with the one provided in the previous section. Fig. 8 illustrates the sorted F1-score across all subjects using iRPF and RPF with five rejection thresholds (0.002, 0.01, 0.05, 0.25, 0.5). As expected, for the RPF, no specific rejection threshold performs better than the others in practice, which highlights the importance of automatic thresholding.

*3.3. Different Choices of Potato Fields*

In this subsection, different designs of the potato field are investigated for the iRPF method to assess how strongly the performance of iRPF is influenced by the specific design employed for the potato field. Fig. 9 illustrate the evaluation metrics for different potato fields in the EYEAR database with and without using EOG channels. As shown, different designs of the potato field, as long as they are properly defined, do not significantly impact the evaluation metrics.

*3.4. Ablation Study*

To assess the contribution of each improvement in the proposed method, we conducted an ablation study by systematically removing each component from the model. This allowed us to evaluate the impact of each feature on the overall performance. As shown in Table 7, the full model, incorporating all components, achieved the highest F1-scores across all databases. Removing any single component resulted in a decrease in performance, particularly when outlier rejection or automatic thresholding were excluded.

Table 7: Ablation Study

| Outlier Rejection | Distance Metrics | Meta-Combination | Automatic Thresholding | EYEAR with EOG | EYEAR w/o EOG | TUAR |
|---|---|---|---|---|---|---|
| | | | | F1-score | | |
| ✓ | ✓ | ✓ | ✓ | 0.96 ± 0.03 | 0.90 ± 0.05 | 0.42 ± 0.14 |
| ✗ | ✓ | ✓ | ✓ | 0.78 ± 0.14 | 0.69 ± 0.13 | 0.39 ± 0.14 |
| ✗ | ✗ | ✓ | ✓ | 0.73 ± 0.15 | 0.65 ± 0.12 | 0.37 ± 0.15 |
| ✗ | ✗ | ✗ | ✓ | 0.73 ± 0.14 | 0.66 ± 0.12 | 0.36 ± 0.15 |
| ✗ | ✗ | ✗ | ✗ | 0.58 ± 0.19 | 0.57 ± 0.13 | 0.35 ± 0.15 |

In the absence of automatic thresholding, a threshold of 0.01 was adopted, as it yielded the best results. Legend: ✓= component enabled, ✗= component disabled.

## 4. Discussion

The improved Riemannian Potato Field method presented in this study holds promise to advance the practice of automated artifact rejection for EEG data, effectively addressing key limitations of traditional Riemannian geometry-based approaches. The analysis demonstrated that iRPF consistently outperforms selected state-of-the-art artifact rejection methods, namely, RPF, RP, IF, and AR, in terms of recall, specificity, precision, and F1-score across a variety of artifact types collected in both clinical and laboratory environments. Statistical and effect size analyses further confirmed the significance of these improvements, highlighting the reliability of iRPF in distinguishing clean EEG segments from artifacts—a critical advantage for EEG-based applications such as BCIs.

From a methodological standpoint, the proposed iRPF method demonstrates superior reliability compared to the selected state-of-the-art techniques. RP employs a single potato encompassing all available electrodes, which results in decreased performance—particularly when artifacts affect only a few channels, as is often the case with eye-related artifacts. In contrast, both iRPF and RPF improve performance by utilizing multiple low-dimensional potatoes. However, RPF's effectiveness is limited by the relatively low accuracy of its derived SQIs and the absence of an automatic, adaptive thresholding mechanism—two limitations that are effectively addressed in the proposed iRPF method. Compared to the multivariate approach employed by Riemannian geometry-based methods, both IF and AR eliminate artifacts using thresholding strategies based on peak-to-peak amplitude values—an approach that may not effectively distinguish between clean and artifact-contaminated epochs, as it does not explicitly account for the characteristic patterns of EEG artifacts. AR, in particular, relies on the Frobenius norm of the mean and median values of the peak-to-peak amplitude. It implicitly assumes that epochs with peak-to-peak values higher than the mean or median are likely to contain artifacts, while those with lower values are considered clean. As a result, low-amplitude eye-related artifacts—such as HEM and VEM or weak blinks—may go undetected. Unlike AR, the IF method uses the minimum value as a criterion to distinguish between clean epochs and artifacts during the iterative process. However, this leads to excessive rejection of clean data segments, thereby limiting its practical applicability. Moreover, the performance of the AR method is suboptimal when the proportion of artifact-contaminated segments exceeds that of clean data—a condition observed in both databases used in this study. Furthermore, in terms of execution time as reported in Table 8, iRPF performed comparably to RPF and IF, while significantly outperforming AR, which exhibits the longest processing times due to its reliance on Bayesian optimization. The efficiency of iRPF makes it highly suitable for large-scale EEG studies and real-time applications, where rapid artifact rejection is crucial to preserving signal quality with minimal delay.

Table 8: Total Execution Time and Execution Time Per Epoch (in Seconds)

| Database | iRPF | | RPF | | RP | | IF | | AR | |
|---|---|---|---|---|---|---|---|---|---|---|
| | Epoch | Total | Epoch | Total | Epoch | Total | Epoch | Total | Epoch | Total |
| EYEAR with EOG | 0.005 | 15.78 | 0.006 | 16.83 | 0.037 | 105.96 | 0.005 | 14.7 | 0.381 | 1072.7 |
| EYEAR w/o EOG | 0.008 | 22.46 | 0.008 | 23.79 | 0.034 | 96.19 | 0.005 | 14.69 | 0.339 | 956.37 |
| TUAR | 0.003 | 180.65 | 0.007 | 427.02 | 0.008 | 495.35 | 0.003 | 141.47 | 0.106 | 6450.89 |

A key advancement of the iRPF method is its ability to efficiently and automatically determine all required hyperparameters, enhancing its generalizability across different databases. As demonstrated in comparisons with the RPF, the proposed iRPF method consistently achieved higher F1-scores by adaptively fine-tuning its rejection criteria, highlighting the limitations of manually fixed rejection thresholds. We do not claim that the iRPF method can always determine the appropriate rejection threshold for every database under all circumstances. Rather, while awaiting further improvements and validations, we suggest that it can be used as an aid for the gold-standard manual artifact rejection method; instead of manually inspecting and rejecting segments throughout the recording, iRPF can display epochs sorted by SQI, allowing a human expert to simply verify the appropriateness of the threshold set by iRPF, and adjust it if necessary. As revealed by the ablation study, a key advancement of the method is its ability to automatically remove outliers, which can significantly affect barycenter estimation and decrease the performance of Riemannian geometry-based approaches.

While iRPF provides robust and adaptive artifact rejection, certain limitations persist. The limited availability of publicly accessible EEG databases annotated for artifacts constrained our evaluation to only two databases, each presenting specific limitations. The EYEAR database contains solely eye-related artifacts and therefore lacks diversity in artifact types. Additionally, in certain EEG files of the TUAR database, the proportion of clean data is insufficient to effectively train the artifact rejection methods. This reminds that artifact rejection approaches such as iRPF need a sufficient amount of clean signal, in general at least 30 seconds. We have found that variations in the design of the potato field do not significantly impact the method's performance (data not shown); however, a comprehensive understanding of artifact types, their physiological origins, frequency characteristics, and associated electrodes is still necessary for the optimal design of the potato field. Another limitation is that iRPF does not currently offer a specific solution for artifacts affecting only a single channel (channel selection), which may require additional pre-processing steps or integration with other channel noise cancellation techniques.

Future research could further enhance iRPF through several avenues. One promising direction is the additional validation of iRPF across a broader range of EEG devices and applications, which would provide a more comprehensive evaluation of its generalizability and efficiency. An interesting avenue for future research is the development of a tool that can define an appropriate potato field based on the available EEG electrodes. The iRPF method is currently limited to offline artifact rejection, restricting its real-time application, particularly in EEG-based BCIs. While the improvements in iRPF have reduced the computational cost compared to existing offline methods, further development is needed to enable its use in online settings. Another avenue involves the classification of artifacts using alternative distance metrics explored in this study, which may enhance the differentiation of artifact types and enable more targeted rejection strategies. Although several studies have assessed the impact of artifact correction in the context of deep learning-based BCIs, artifact rejection techniques remain relatively underexplored. This underscores the need for systematic investigations into how the removal of contaminated segments may affect the robustness and generalizability of deep learning models.

**CRediT authorship contribution statement**

**Davoud Hajhassani:** Conceptualization, Methodology, Software, Validation, Formal analysis, Investigation, Data curation, Writing – original draft, Writing – review & editing, Visualization. **Quentin Barthélemy:** Conceptualization, Methodology, Writing – review & editing. **Jérémie Mattout:** Methodology, Resources, Writing – review & editing, Supervision, Project administration, Funding acquisition. **Marco Congedo:** Conceptualization, Methodology, Software, Formal analysis, Investigation, Resources, Writing – review & editing, Supervision, Project administration.

**Declaration of interest**

None of the authors have potential conflicts of interest to be disclosed.

**Acknowledgement**

This work has been supported by ANR under HiFi grant ANR-20-CE17-0023. The authors would like to thank the anonymous reviewers for their constructive suggestions, which significantly contributed to improving the original manuscript.

**Ethical Consideration**

Open-source databases were used in this study; therefore, ethical approval was waived, as the data were anonymized and did not involve direct interaction with human participants.

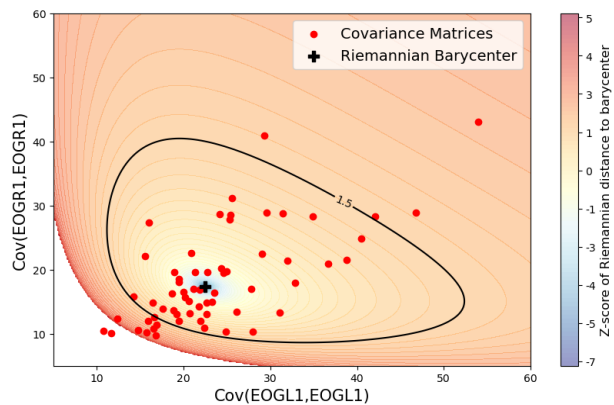

**Fig. 1:** 2D projection of the z-score map for a Riemannian potato, shown for two selected channels (EOGL1, EOGR1) across 66 epochs from an EYEAR database (R. Kobler et al., 2017), along with their Riemannian barycenter. The colormap represents z-scores, with a selected isocontour outlining the acceptance region of the potato ($z_{th} = 1.5$ to enhance visualization). Points outside the potato are considered artifacts and are therefore rejected.

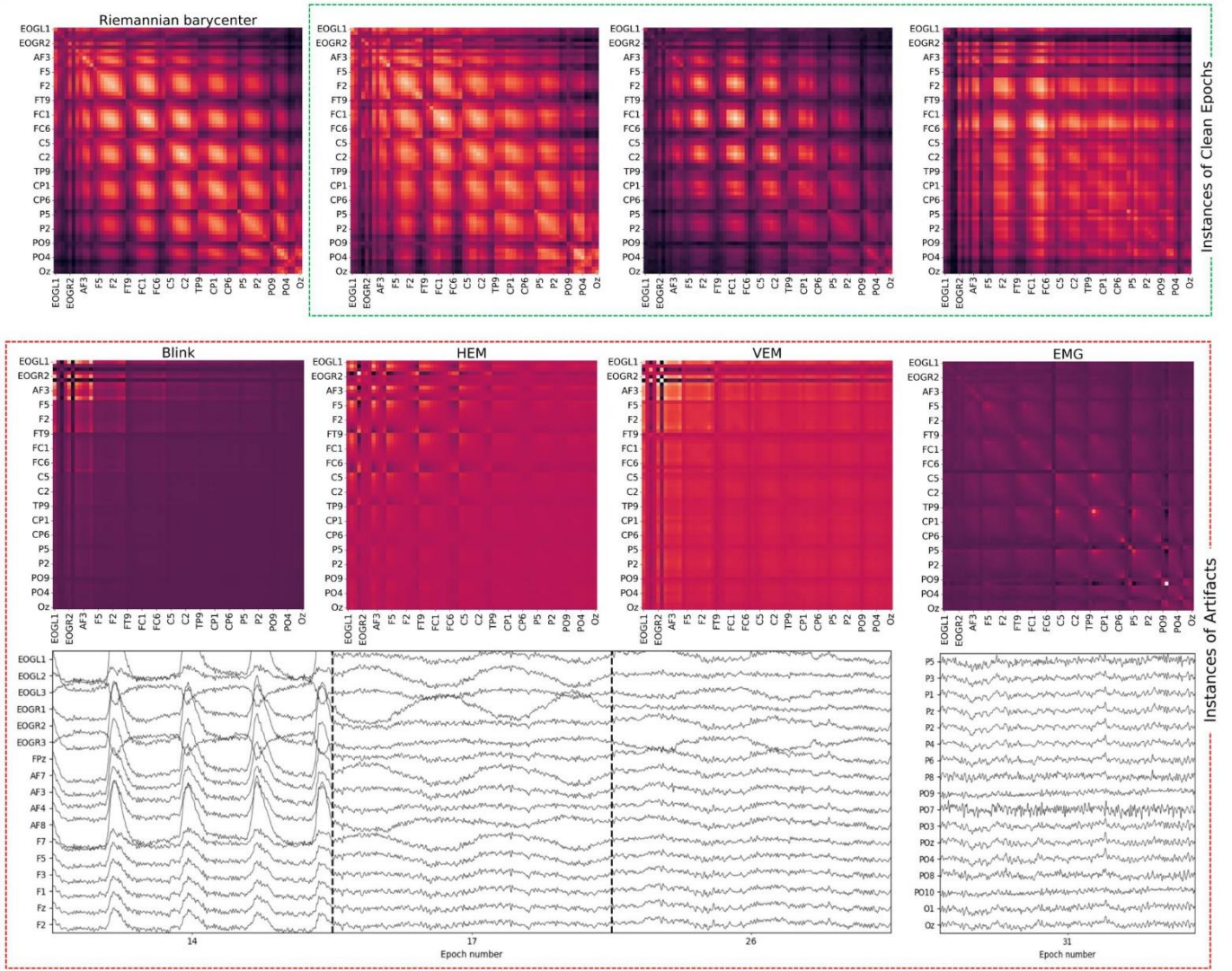

**Fig. 2:** Comparison of heatmaps depicting the Riemannian barycenter estimated on clean epochs, along with covariance matrices of single instances of clean epochs (first row) and epochs contaminated (second row) by various types of artifacts, including blinks, horizontal eye movements (HEM), vertical eye movements (VEM), and EMG. For each artifact, the corresponding EEG data is displayed below the heatmap. Each heatmap is scaled to its own absolute maximum. As it can be appreciated from the heatmaps, the covariance matrices of clean epochs closely resemble the covariance matrix of the barycenter. This is notably not the case for epochs with artifact contamination, which diverge significantly. This divergence is precisely what the affine-invariant Riemannian distance in Eq. (2) measures. Heatmaps are plotted for blink, HEM, and VEM in the frequency band of 0.1 to 8 Hz, and for EMG in the range of 20 to 44 Hz. The data is from subject 20 of the EYEAR database [84].

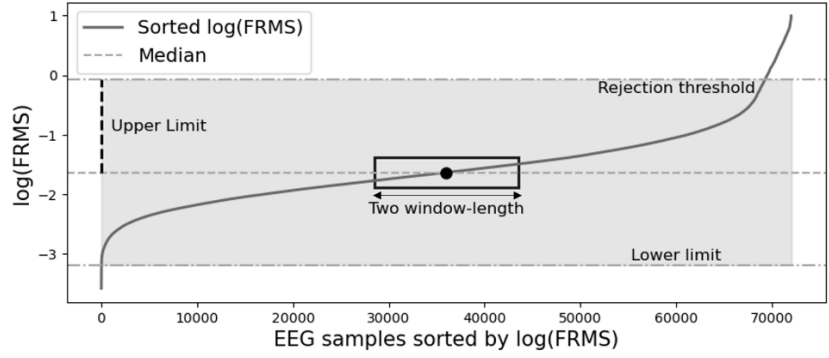

**Fig. 3:** Illustration of the proposed EEG outlier rejection method. The rectangle indicates the window twice the window-length, centered around the median, used to compute the mean FRMS (black dot). Trials are rejected if any of their samples exceed the rejection threshold.

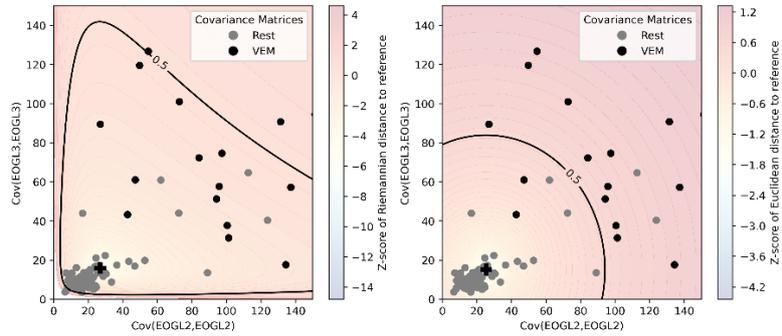

**Fig. 4:** Illustration of Euclidean distance (**right**) versus Riemannian distance (**left**) in distinguishing the covariance matrices of epochs labeled as vertical eye movement (VEM) from the covariance matrices labeled as resting state (rest). The Riemannian barycenter is represented by the black ✚ sign. The colormap indicates the z-score, and a selected isocontour ($z_{th} = 0.5$ to enhance visualization) defines the boundary of the potato. As observed, the Euclidean distance more effectively separates vertical eye movements from resting state epochs compared to the Riemannian distance. This analysis is performed using data from Subject 5 of the EEG eye artifact database [84], considering EOGL2 and EOGL3 as the electrodes defining the potato. The signal is band-pass filtered between 0.1 to 7 Hz and only epochs labeled as resting state or vertical eye movement are included.

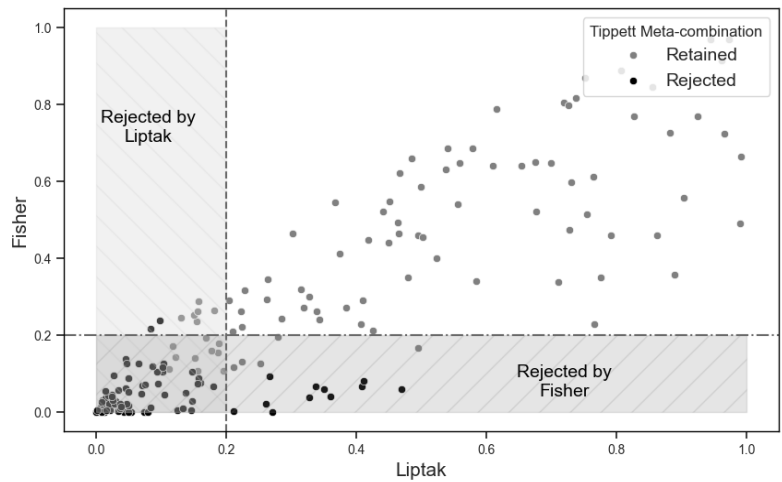

**Fig. 5:** Application of various p-value combination functions and a meta-combination function in the determination of the rejection region. Using a threshold of 0.2, the rejection regions for the Fisher method in the RPF approach and the Liptak method are highlighted. The rejection region in the proposed iRPF method, determined by applying the Tippett meta-combination function to the Liptak and Fisher methods, is shown with the rejected points. The p-values are derived from Subject 20 of the EEG eye artifact database [84].

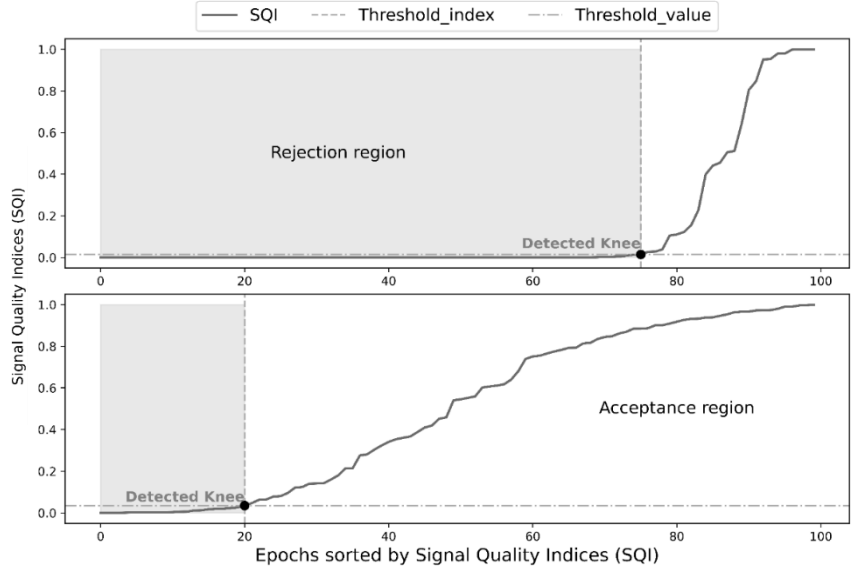

**Fig. 6:** Automatic thresholding for identifying rejection and acceptance (clean epochs) regions by applying the knee detection algorithm on sorted Signal Quality Index (SQI) values for the iRPF method (**Top**) and RPF method (**Bottom**). The SQIs are derived from Subject 20 of the EEG eye artifact database [84].

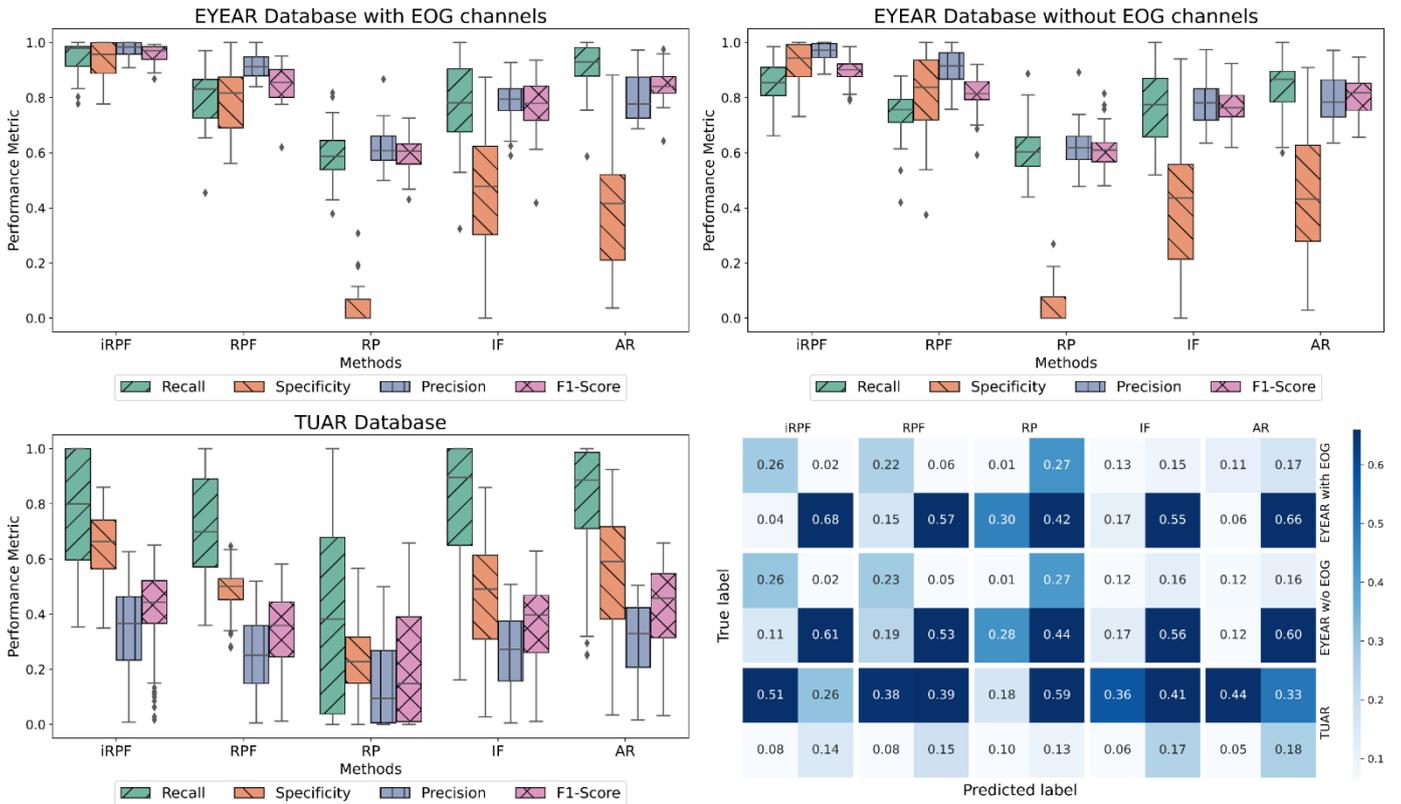

**Fig. 7:** Comparison of Recall, Specificity, Precision, and F1-Score, averaged across all subjects for our proposed method and other artifact rejection methods. The normalized confusion matrices are also provided for each method across all databases. **iRPF**, improved Riemannian Potato Field; **RPF**, Riemannian Potato Field with 0.5 as a threshold; **RP**, Riemannian Potato with 2.0 as a threshold; **IF**, Isolation Forest; **AR**, Autoreject.

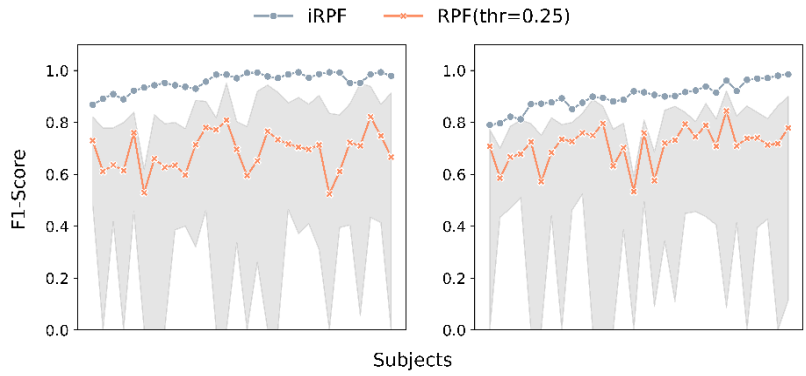

**Fig. 8:** Comparison of F1-Score for the EYEAR database, with (**Left**) and without (**Right**) using EOG channels, between **iRPF** and **RPF** with different thresholds (0.002, 0.01, 0.05, 0.25, 0.5).

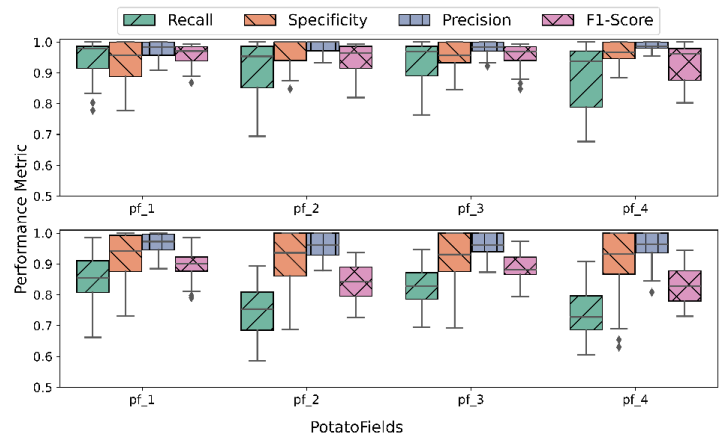

**Fig. 9:** Comparison of Recall, Precision, Specificity, and F1-Score, averaged across all subjects, for the iRPF method applied to the EYEAR database. This comparison is made across different definitions of potato fields with (**Top**) and without (**Bottom**) using EOG channels.